\documentstyle[aps,prl]{revtex}

\begin{document}

\draft
\title{
Noise Suppression by Noise
}
\author{
J. M. G. Vilar$^1$
and J. M. Rub\'{\i}$^{2}$
}
\address{ 
$^1$Departments of Physics and Molecular Biology,
Princeton University, Princeton, New Jersey 08544\\
$^2$Departament de F\'{\i}sica Fonamental, Facultat de
F\'{\i}sica, Universitat de Barcelona,
Diagonal 647, E-08028 Barcelona, Spain
}
\maketitle
\begin{abstract}
We have analyzed the interplay between an externally added noise and
the intrinsic noise of systems that relax fast towards
a stationary state,
and found that increasing the intensity of the external noise
can reduce the total noise of the system.
We have established a general criterion for the appearance of
this  phenomenon and discussed two examples in detail.
\end{abstract}
\pacs{PACS numbers: 
05.40.-a }

For a long time, noise was considered to be only a source of disorder,
a nuisance to be avoided. Recently, this view has been changing due to
several phenomena that show constructive facets
of noise. Among them,
the most widely studied is the phenomenon of stochastic
resonance, where the addition of noise to a system enhances its response
to a periodic force~\cite{first,revs}. This counterintuitive aspect of
noise has been found
under a wide variety of situations, including: bistable~\cite{bi} and
monostable~\cite{mono} systems; nondynamical elements with~\cite{withre}
and without~\cite{wothre} threshold; and pattern forming
systems~\cite{patt}. Similar constructive outcomes
are also found in other remarkable phenomena, such as
noise induced transitions~\cite{HoLe} and noise induced transport~\cite{JP}.
To some extent,  the presence of noise is an unavoidable
feature,  and as one moves
from macroscopic to microscopic scales that presence becomes more and more
prominent. To withdraw the noise,
it is customary to reduce as much as possible all the external
noise sources that affect the system since it still seems
paradoxical that adding noise might result in a
less noisy system.

In this Letter we show that the intrinsic
noise displayed
by some systems can substantially be reduced through its nonlinear
interplay  with  externally added noise and we
establish sufficient conditions for this phenomenon to occur.
The systems we consider are those relaxing fast to a stationary state where
their properties are completely determined by some parameters:
the state of the system, denoted  by $I(t, {\bf V})$,
is a function of some input parameters, denoted by the set $\bf V$.
These systems are usually called nondynamical systems. 
For the sake of simplicity, we consider the case of a single input
parameter, i.e. ${\bf V} \equiv V$. The temporal dependence of the
state of the system takes into account the intrinsic fluctuations.
This stochastic behavior can be described
by the mean value
\begin{equation}
\left<I(t,V)\right>=H(V)
\end{equation}
and the correlation function
\begin{equation}
\left<I(t,V)I(t+\tau,V)\right>-\left<I(t,V)\right>^2=\tilde G(V,\tau) \;,
\end{equation}
where $\langle . \rangle$ indicates average over the noise.
For practical purposes, it is convenient to write $I(V,t)$ as
\begin{equation}
I(t,V)=H(V)+\xi(t,V)\;,  \label{master-1}
\end{equation}
where $\xi(t,V)$, with $\left<\xi(t,V)\xi(t+\tau,V)\right>=\tilde G(V,\tau)$,
represents the intrinsic noise.
When the characteristic time scales of the fluctuations are smaller
than any other entering the system,
$\tilde G(V,\tau)=G(V)\delta(\tau) \;,$
and the power spectrum of the fluctuations
is flat for the range of frequencies of interest.
The function $H(V)$ corresponds to the deterministic response
and $G(V)$ to the intensity of the intrinsic fluctuations.

We now analyze how adding external noise affects the output of
the system. We consider that a
random quantity $\zeta (t)$ is added to a constant input $V_{0}$; i.e.,
$V(t)=V_{0}+\zeta (t)$. The random term is assumed to be Gaussian noise with
zero mean and correlation function $\left\langle \zeta (t)\zeta (t+\tau
)\right\rangle =\sigma ^{2}\exp (-\tau /\tau _{F})$, where $\sigma ^{2}$
defines the noise level and $\tau _{F}$ is the correlation time. If the
correlation time $\tau _{F}$ is not too small,  the system
adapts to each value of $V(t)$ and 
can be described by
an input-output relationship where the input is $V_{0}$.
The external noise, however, changes the characteristics of the system
which is then described by 
\begin{equation}
I(t,V_0)=H_{0}(V_{0})+\xi _{0}(t,V_0)\;,  \label{master0}
\end{equation}
where the mean value and the total fluctuations of the output are taken into
account by the terms $H_{0}(V_{0})$ and $\xi _{0}(t,V_0)$, respectively.
For times
higher than $\tau_F$, the
noise term can again be approximated by Gaussian white noise with zero mean,
but now with correlation function
$\left\langle \xi _{0}(t,V_0)\xi _{0}(t+\tau,V_0)
\right\rangle =G_{0}(V_0)\delta (\tau )$.

For small $\sigma ^{2}$,
$H_{0}(V_0)=\left\langle I[t,V(t)]\right\rangle $
and $G_{0}(V_0)=\int_{-\infty }^{\infty }[\left\langle I[t,V(t)]I[t+\tau,
V(t+\tau)]\right\rangle
-\left\langle I[t,V(t)]\right\rangle ^{2}]d\tau$ can
readily be obtained by expanding $H(V)$ and $G(V)$ in power series of
$V$ around $V_{0}$.
Now the average $\left\langle .\right\rangle $ must be carried
out over the two noises. By averaging first over the internal noise,
we obtain
\begin{eqnarray}
H_{0}(V_{0}) &=&H(V_0)
+{\frac{1}{2}}{H^{\prime \prime }}(V_{0})\left<\zeta (t)^2\right>\;, \\
G_{0}(V_{0}) &=&\int_{-\infty }^{\infty }
\left\{\left[G(V_{0})+\frac{1}{2}
{G^{\prime \prime }}(V_{0})\left<\zeta (t)^2\right>\right]\delta(\tau)
+{H^{\prime }}(V_{0})^{2}\left<\zeta (t)\zeta (t+\tau)\right>\right\}d\tau
\;, 
\end{eqnarray}
where $^{\prime}$ indicates the derivative of the function with respect to
its argument.
The second average leads to 
\begin{eqnarray}
H_{0}(V_{0}) &=&H(V_{0})
+{\frac{1}{2}}{H^{\prime \prime }}(V_{0})\sigma ^{2}\;, \\
G_{0}(V_{0}) &=&G(V_{0})+\left[2\tau _{F}{H^{\prime }}(V_{0})^{2}+{\frac{1}{2}}
{G^{\prime \prime }}(V_{0})\right]\sigma ^{2}\;. \label{condition1}
\end{eqnarray}
 Therefore,
the output noise may be decreased by the addition of external noise when
$G^{\prime \prime }(V_{0})$ is negative and the correlation time of the
external noise is sufficiently small.

It may also be interesting to decrease the output
noise intensity for a fixed mean value of the output. For this purpose,
the input 
$V_0$ must be tuned to a new value $V_{c}$
so that 
\begin{eqnarray}
H_{0}(V_{c}) &=&H(V_{0})\;, \\
G_{0}(V_{c}) &=&G(V_{0})+\Delta G(V_{0})\;.
\end{eqnarray}
If the noise level is small, the new value of the input is 
\begin{equation}
V_{c}=V_{0}-{\frac{{H^{\prime \prime }}(V_{0})}{2H^{\prime }(V_{0})}}\sigma
^{2}\;,
\end{equation}
and the variation of the output noise intensity $\Delta G(V_{0})$ is given
by 
\begin{eqnarray}
{\frac{\Delta G(V_{0})}{\sigma ^{2}}} &=&2\tau _{F}{H^{\prime }}(V_{0})^{2}+
{\frac{1}{2}}{G^{\prime \prime }}(V_{0})  
-G^{\prime }(V_{0}){{H^{\prime \prime }}(V_{0})/2H^{\prime }(V_{0})}\;,
\label{condition2}
\end{eqnarray}
which can take negative values as well. 

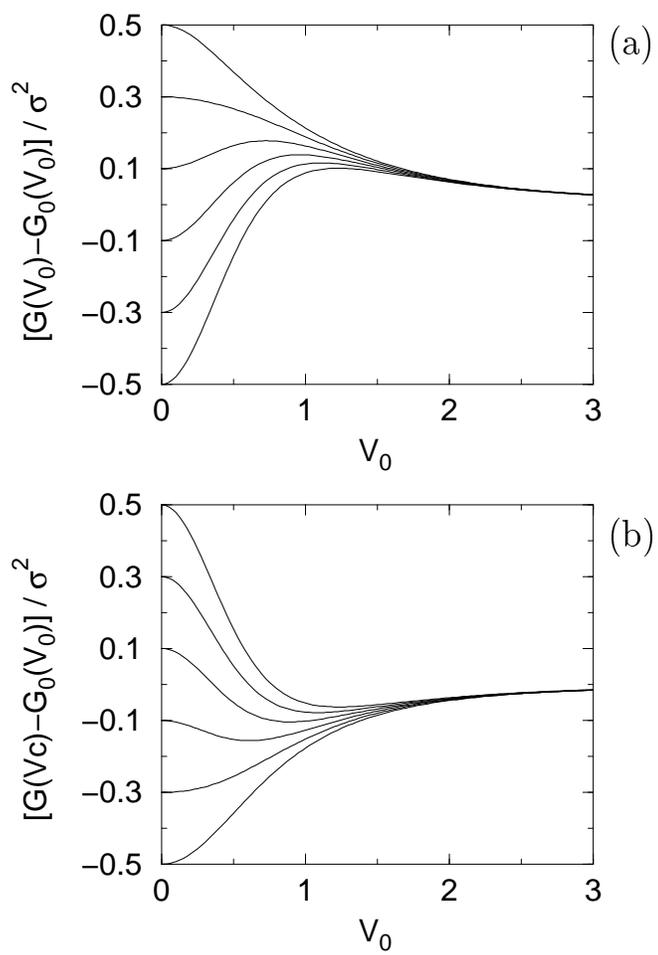
\begin{figure}[th]
\caption[a]{\label{fig1}
Increase of the output noise over the input
noise for the nonlinear electric conduction model
[Eqs. (\ref{modelo11})
and (\ref{modelo12})] with $Q=1$ and $R=1$.
The different lines correspond to different
input noise correlation times (from bottom to top):
$\tau_F=0.0$, $0.1$, $0.2$, $0.3$, $0.4$, $0.5$.
All quantities
and parameters are given in arbitrary units.
}
\end{figure}

As a first example illustrating the applicability of our results, we will
analyze a model for electrical conduction which displays saturation. In this
model, $I(t,V)$ corresponds to the current intensity and $V$ to an input
voltage. To be explicit, we consider 
\begin{eqnarray}
H(V) &=&{\frac{V}{R(1+V^{2})^{1/2}}}\;\;,  \label{modelo11} \\
G(V) &=&{\frac{Q}{(1+V^{2})^{1/2}}}\;\;, \label{modelo12} 
\end{eqnarray}
where $R$ and $Q$ are constants.
A well known example of systems exhibiting this non-linear 
behavior are semiconductor systems that display hot electrons
effects~\cite{hot2,hot1}. 
According to our previous analysis, the output noise for an input with
mean value $V_{0}$ is given by 
\begin{eqnarray}
G(V_{0}) &=&G_{0}(V_{0}) +\,{\frac{4\tau _{F}-QR^{2}\left( 1-2V_{0}^{2}\right)
(1+V_{0}^{2})^{1/2}}{
2R^{2}\left( 1+V_{0}^{2}\right) ^{3}}}\,\sigma ^{2}\;.
\end{eqnarray}
This indicates that for small correlation times and small mean voltages,
the output noise of the device may be decreased by the addition of noise.
Similar results are obtained when the input $V_{0}$ is
changed to $V_{c}$ in order for the mean value of the output to have the same
value as in absence of noise. For the new value of the input,
\begin{equation}
V_{c}=V_{0}+{\frac{3V_{0}}{2+2V_{0}^{2}}}\,\sigma ^{2}\;,
\end{equation}
the output noise changes to 
\begin{equation}
G(V_{c})=G_{0}(V_{0})+{\frac{4\tau _{F}-QR^{2}\left( 1+V_{0}^{2}\right)
^{3/2}}{2R^{2}\left( 1+V_{0}^{2}\right) ^{3}}}\,\sigma ^{2}\;.
\end{equation}
As in the previous case, the output noise can be decreased for sufficiently
small correlation time of the input noise, but now, it is not required for
the applied voltage to be small. These features are illustrated in Fig.
\ref{fig1} for different values of the correlation time $\tau_{F}$.

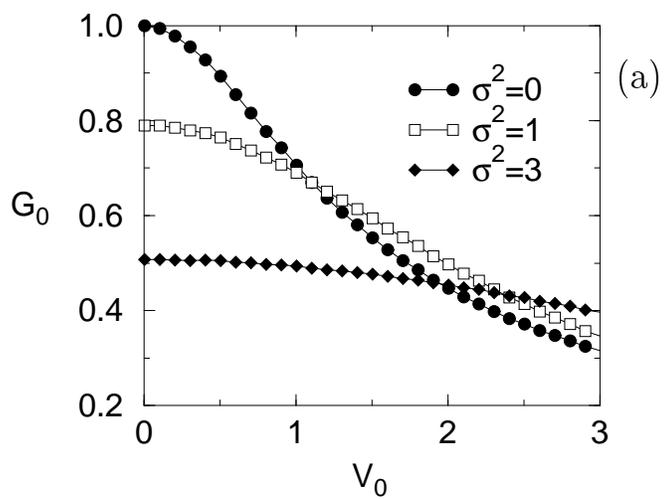
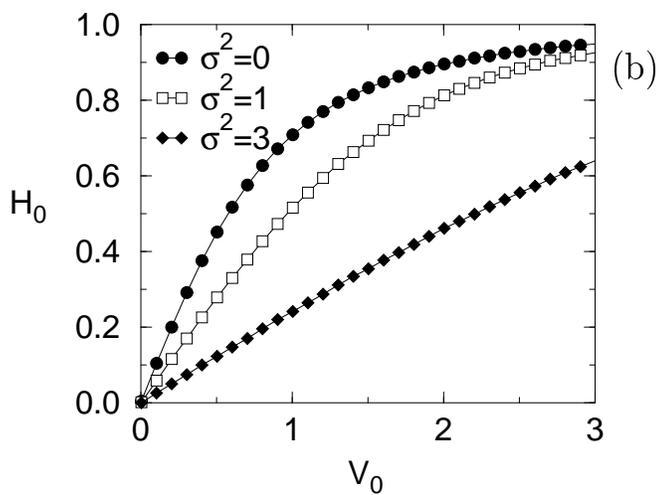
\begin{figure}[th]
\caption[a]{\label{fig2}
Change of (a) the output mean and (b) the
output fluctuations  for the nonlinear electric
conduction model [Eqs. (\ref{modelo11})
and (\ref{modelo12})] with $\tau_F=0.001$, $Q=1$, and $R=1$.
All quantities
and parameters are given in arbitrary units.
}
\end{figure}

To extend our results to higher input noise levels, we have computed
$H_0(V_0)$ and $G_0(V_0)$ by numerically averaging
Eqs. (\ref{modelo11}) and (\ref{modelo12}) over the fluctuating input.
In Fig. \ref{fig2} we have shown the results for different
intensities of the noise. This figure indicates that qualitatively the same
phenomenon appears for higher noise intensity.

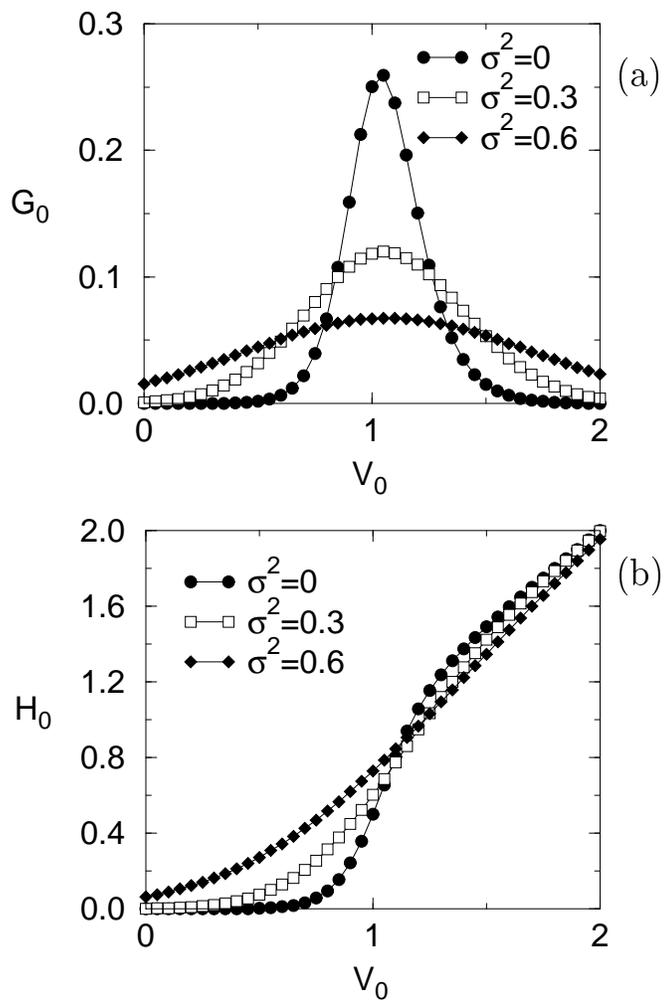
\begin{figure}[th]
\caption[a]{\label{fig3}
Change of (a) the output mean and (b) the
output fluctuations  for the ionic channel
model [Eqs. (\ref{modelo21})
and (\ref{modelo22})] with $\tau_F=0.001$,
$\Delta=10$, and $W=1$.
All quantities
and parameters are given in arbitrary units.
}
\end{figure}

The second example we consider is an ionic channel model~\cite{chan}.
The characteristics of the current intensity $I(t,V)$ as a
function of the input voltage are given by
\begin{eqnarray}
H(V) &=&{\frac{V}{1+e^{\Delta(W-V)}}}\;\;,  \label{modelo21} \\
G(V) &=&{\frac{V^2e^{\Delta(W-V)}}{\left(1+e^{\Delta(W-V)}\right)^2}}\;\;, 
\label{modelo22} 
\end{eqnarray}
where $W$ and $\Delta$ are constants. By using Eqs. (\ref{condition1}) and
(\ref{condition2}) one can  easily see that the input noise can reduce the
output noise for a fixed average of either the input or the output.
Fig. \ref{fig3} shows the results obtained by numerically averaging
Eqs. (\ref{modelo21}) and (\ref{modelo22}). As in the previous example,
there is a range of values of $V_0$ where the input noise can reduce the
output noise.

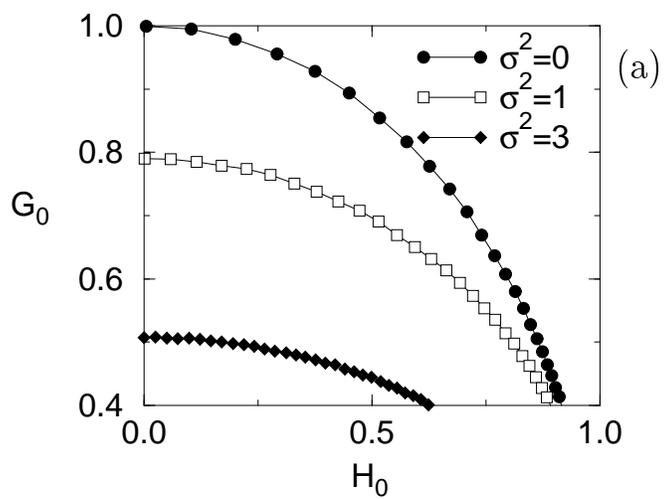
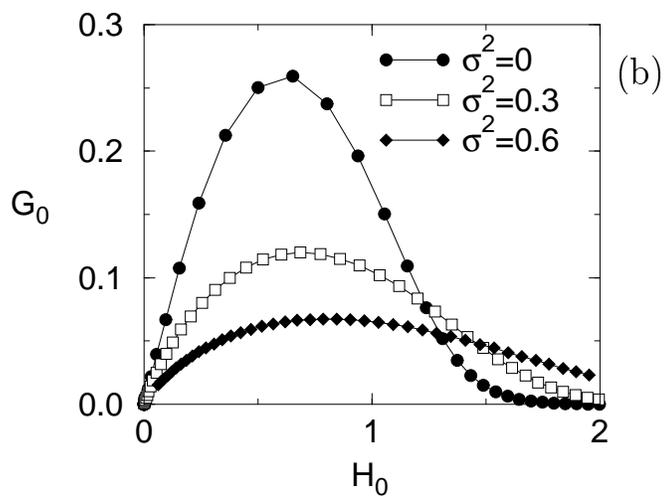
\begin{figure}[th]
\caption[a]{\label{fig4}
Output noise intensity as a function of the mean output
for (a) the nonlinear conduction model [Eqs. (\ref{modelo11})
and (\ref{modelo12}) with $\tau_F=0.001$, $Q=1$, and $R=1$]  
and (b) the ionic channel model
[Eqs. (\ref{modelo21}) and (\ref{modelo22})
with $\tau_F=0.001$, $\Delta=10$, and $W=1$]. All quantities
and parameters are given in arbitrary units.
}\end{figure}

The way noise affects the output of the two systems considered previously
is depicted in Fig. \ref{fig4}. This figure shows the output noise intensity
as function of the mean value of the output. For the values of the
parameters used in Fig. \ref{fig4}(a), 
the system can have the same average output with less
fluctuations by just adding
noise. Similarly, in  Fig. \ref{fig4}(b), externally added
noise greatly reduces the output noise around $H_0=0.5$, but, for higher
values of $H_0$, the situation
is just the opposite: noise plays its usual
detrimental role.

Our analysis indicates that addition of noise to systems
that have intrinsic noise can change their properties to the extent 
that they may display
less noise. Other studies have shown
that intrinsic noise can be responsible 
for the appearance of stochastic resonance~\cite{withre,nnd1} and
aperiodic stochastic
resonance~\cite{nnd2} in nondynamical systems. In those studies,
external noise enhances
the response of the system to a periodic or an aperiodic signal. In our case,
noise does not need to cooperate with an external signal to play a
constructive role. A line of investigation for future work would be to
extend our results to dynamical systems.
This would need the development of new methodologies, in a similar way as
has been done for nonlinear noise sources in stochastic differential
equations~\cite{HoLe,bnas}. Our work then offers new perspectives on
what concerns the constructive effects of noise in general nonlinear systems.

This work was partially supported by DGICYT (Spain) Grant No.
PB98-1258. J.M.G.V. wishes to thank MEC (Spain) for financial
support.

\begin{references}

\bibitem{first}  R. Benzi, A. Sutera, and A. Vulpiani, J. Phys. A {\bf 14},
L453 (1981); S. Fauve and F. Heslot, Phys. Lett. A {\bf 97}, 5 (1983);
B. McNamara, K. Wiesenfeld, and R. Roy, Phys. Rev. Lett. {\bf 60},
2626 (1988); L. Gammaitoni, F. Marchesoni, E. Menichella-Saetta, and S.
Santuchi, Phys. Rev. Lett. {\bf 62}, 2626 (1989).

\bibitem{revs} F. Moss, in {\it Contemporary Problems
in Statistical Physics},
edited by G. H. Weiss (SIAM, Philadelphia,1994);
K. Wiesenfeld and F. Moss, Nature {\bf 373}, 33 (1995);
L. Gammaitoni, P. H\"anggi, P. Jung, and F. Marchesoni,
Rev. Mod. Phys. {\bf 70}, 223 (1998).

\bibitem{bi}  B. McNamara and K. Wiesenfeld,  Phys. Rev. A {\bf 39},
4854 (1989).

\bibitem{mono}  J. M. G. Vilar and J. M. Rub\'{\i }, Phys. Rev. Lett.
{\bf 77}, 2863 (1996). 

\bibitem{withre}  Z. Gingl, L. B. Kiss, and F. Moss, Europhys. Lett. {\bf 29}
, 191 (1995); F. Chapeau-Blondeau and X. Godivier, Phys. Rev. E {\bf 55},
1748 (1997).

\bibitem{wothre}  S. M. Bezrukov and I. Vodyanoy, Nature {\bf 385}, 319 (1997).

\bibitem{patt} P. Jung and G. Mayer-Kress, Phys. Rev. Lett. {\bf 74},
2130 (1995); J. M. G. Vilar and J. M. Rub\'{\i }, Phys. Rev. Lett.
{\bf 78}, 2886 (1997).

\bibitem{HoLe} W. Horsthemke and R. Lefever,
{\it Noise-Induced Transitions} (Springer, Berlin, 1984).

\bibitem{JP} F. J\"ulicher, A. Ajdari, and J. Prost ,
Rev. Mod. Phys.  {\bf 69}, 1269 (1997).


\bibitem{nnd1} J. M. G. Vilar, G. Gomila, and J. M. Rub\'{\i}, 
Phys. Rev. Lett. {\bf 81}, 14 (1998).

\bibitem{nnd2} P. C. Gailey, A. Neiman, J. J. Collins, and F. Moss,
Phys. Rev. Lett. {\bf 79}, 4701 (1997).

\bibitem{hot2}  L. Reggiani (Ed.), {\it Hot Electron Transport in
Semiconductors} (Springer-Verlag, Berlin, 1985).

\bibitem{hot1}  A. van der Ziel, {\it Noise in Solid State Devices and
Circuits} (John Wiley \& Sons, New York, 1986).

\bibitem{chan} L. J. DeFelice, {\it Introduction to Membrane Noise}
(Plenum Press, New York, London, 1981). 

\bibitem{bnas}  M. San Miguel and J. M. Sancho,
Z. Phys. B {\bf 43}, 361 (1981); F. Sagues, M. San Miguel,
and J. M. Sancho,  Z. Phys. B {\bf 55}, 269 (1984).


\end{references}
\end{document}